\documentclass{INTERSPEECH2023}
\usepackage{bbm}
\usepackage{hyperref}
\hypersetup{hypertex=true,
	colorlinks=true,
	linkcolor=blue,
	anchorcolor=green,
	citecolor=green}
\usepackage{multirow}
 \interspeechcameraready

\title{Harmonic enhancement using learnable comb filter for light-weight full-band speech enhancement model}
\name{ Xiaohuai Le$^1$$^,$$^2$, Tong Lei$^1$$^,$$^3$, Li Chen$^2$, Yiqing Guo$^2$, Chao He$^2$, Cheng Chen$^2$,  \\ Xianjun Xia$^2$, Hua Gao$^2$, Yijian Xiao$^2$, Piao Ding$^2$, Shenyi Song$^2$, Jing Lu$^1$$^,$$^3$}

\address{
	$^1$Key Laboratory of Modern Acoustics, Nanjing University, Nanjing 210093, China\\
	$^2$RTC Lab, ByteDance, China \\
	$^3$NJU-Horizon Intelligent Audio Lab, Horizon Robotics, Beijing 100094, China}
\email{\{xiaohuaile, tonglei\}@smail.nju.edu.cn, \{chenli.cloud, guoyiqing, hechao.12345678, chencheng.c, gaohua.karl, lindanfeng, dingpiao, songshenyi\}@bytedance.com, xxjpjj@mail.ustc.edu.cn, lujing@nju.edu.cn}

\begin{document}
	
\maketitle

\begin{abstract}
With fewer feature dimensions, filter banks are often used in light-weight full-band speech enhancement models. In order to further enhance the coarse speech in the sub-band domain, it is necessary to apply a post-filtering for harmonic retrieval. The signal processing-based comb filters used in RNNoise and PercepNet have limited performance and may cause speech quality degradation due to inaccurate fundamental frequency estimation. To tackle this problem, we propose a learnable comb filter to enhance harmonics. Based on the sub-band model, we design a DNN-based fundamental frequency estimator to estimate the discrete fundamental frequencies and a comb filter for harmonic enhancement, which are trained via an end-to-end pattern. The experiments show the advantages of our proposed method over PecepNet and DeepFilterNet.
\end{abstract}
\noindent\textbf{Index Terms}: Comb filter, Speech enhancement, PercepNet, DeepFilterNet

\section{Introduction}

Due to the development of high-quality real-time communication, DNN-based full-band speech enhancement (SE) algorithms have attracted extensive attention \cite{ref1} in both academic researches and industrial applications. Compared with wide-band SE models \cite{ref2}, full-band models are inevitably more complex due to the higher feature dimension.

Besides the straightforward way to design a light-weight model, it is common to use pruning \cite{ref3}, quantization \cite{ref4}, grouping \cite{ref5} and skipping \cite{ref6} techniques to reduce the computational complexity. Multi-stage learning \cite{ref7} is also a popular method for full-band spectrogram processing. As for the feature compression in full-band SE, the most widely used method is applying filter banks taking human perception into account. However, the loss of spectrum details (e.g. the phase and harmonics) caused by the compression limits the performance and results in deteriorated enhanced speech, so that a dedicated post-processing is necessary in order to further suppress the residual noise and improve speech quality \cite{ref8}.

Harmonic enhancement is a useful post-processing method that removes the residual noise between harmonics. RNNoise \cite{ref9} introduced a comb filter to strengthen the harmonics, where a heuristic algorithm is used to estimate the filtering strength. The filtering strength represents the proportion of harmonic-enhanced spectrum. PercepNet \cite{ref10} further improves the comb-filtering algorithm by increasing the denoising depth in the stopbands and using a DNN to estimate the filtering strength. However, the labels of the filtering strength are still derived from harmonic coherence and cannot be optimized end-to-end. Under a multi-task training framework, DeepFilterNet \cite{ref11} uses a convolutional filter called DeepFiltering \cite{ref12} in the time-frequency domain and the coefficients are learned end-to-end without explicit fundamental frequency (F0) information. In Harmonic Gated Compensation Network (HGCN) \cite{ref13}, both F0 estimation and harmonic enhancement are realized by neural networks. However, the spectral resolution is limited due to the frame length of 32 ms, resulting in inferior harmonic enhancement performance than the comb filters.

To take into account both the spectral resolution and the end-to-end optimization, we propose a DNN-based implementation of comb filter. The candidate fundamental frequencies are discretized and estimated utilizing a neural network. On this basis, we use a convolutional neural network (CNN) to realize the comb filtering algorithm. Experiments on a high-performance speech enhancement model, dual-path convolutional recurrent network (DPCRN) \cite{ref14}, show that the proposed method estimates the filtering strength more accurately than PercepNet. With the same latency, the proposed model shows advantages over PercepNet and DeepFilterNet. Meanwhile, the filtering algorithm is aligned with PercepNet during inference with low computational complexity.

\section{Models}
\subsection{Signal model}
The clean speech in the time domain can be expressed following the signal model in PercepNet:
\begin{equation}
	x\left(t\right)=p\left(t\right)+u\left(t\right) \label{eq1}
\end{equation}
where \(x(t)\) denotes the signal at time step \(t\). The \(p(t)\) and \(u(t)\) respectively represent the periodic and the aperiodic components of speech. The aperiodic component can be estimated by ideal ratio mask (IRM). On the basis of IRM, the comb filtering is applied for periodic component estimation. As PercepNet, we use the following symmetric non-causal comb filter for harmonic enhancement:
\begin{equation}
	W_M\left(Z\right)=\sum_{k=-M}^{M}{w_kz^{-kT}} \label{eq2}
\end{equation}
where \(T\) is the period of the F0. The order of the filter is \(2MT+1\) and in this paper \(M=1\). The \(w_k\) is set as the normalized Hanning window coefficient to reduce the spectral leakage of the stopband.

\subsection{Fundamental frequency estimation}
\begin{figure}[tbh!]
	\centering
	\includegraphics[width=0.45\textwidth]{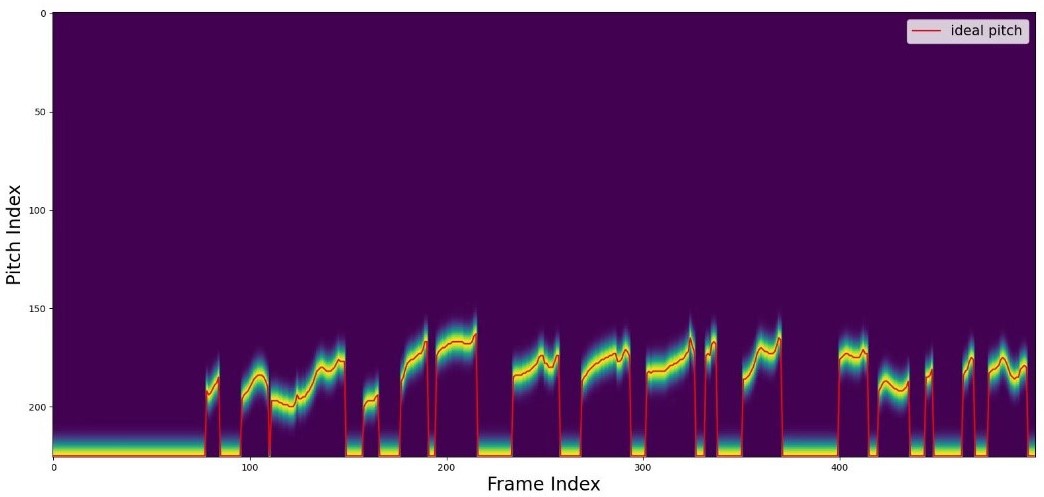}
	\caption{The ideal F0 label. The red line denotes the target F0 and the yellow area is the smoothed label.}
	\label{fig:figure1}
\end{figure}

Accurate fundamental frequency estimation is essential for harmonic enhancement. To enable the end-to-end estimation, we discretize the target fundamental frequency range. Compared with estimating a continuous value of the F0, predicting the discrete frequencies by the multi-classification network is more feasible and there is no significant error if the frequency is sufficiently densely sampled. Similar to CREPE \cite{ref15} and HGCN, the target F0 range is discretized into \(N\) target frequencies. An extra dimension is also added for unvoiced frames. The output of the F0 estimator and the target F0 label at the time step \(t\), are respectively denoted as \(\widehat{\mathbf{f}_t}\) and \(\mathbf{f}_t\), both with dimension \(N+1\). Assuming that the index of a target F0 is \(n\), and then the \(i\)-th term of the Gaussian smoothed F0 label vector \cite{ref15} \(\mathbf{f}_t\) can be calculated by
\begin{equation}
	\mathbf{f}_t\left[i\right]=\text{exp}\left(-\left(i-n\right)^2/50\right)  \label{eq3}                    
\end{equation}
An example of the Gaussian smoothed label is shown in Figure~\ref{fig:figure1}. Compared with the one-hot label, the above label is smoother for training. The target F0 are obtained from pYIN \cite{ref16}, which adopts autocorrelation function and HMM and achieves excellent performance on clean speech. The binary cross entropy (BCE) is used to train the F0 estimator, which can be expressed as
\begin{equation}
	L_{pitch}=-\sum_{i}{\left(\mathbf{f}_t\left[i\right]\text{log}\widehat{\mathbf{f}_t}\left[i\right]+\left(1-\mathbf{f}_t\left[i\right]\right)\text{log}\left(1-\widehat{\mathbf{f}_t}\left[i\right]\right)\right)}
	\label{eq4}
\end{equation}
In this paper, the target F0 range is set as [62.5Hz, 500Hz], and divided into \(N=225\) subintervals by an equal spacing of periods (any reasonable discretization principal such as equipartition of octaves can be used). The 225-th term represents the unvoiced frame.

\subsection{Learnable comb filter}
\begin{figure}[tbh!]
	\centering
	\includegraphics[width=0.5\textwidth]{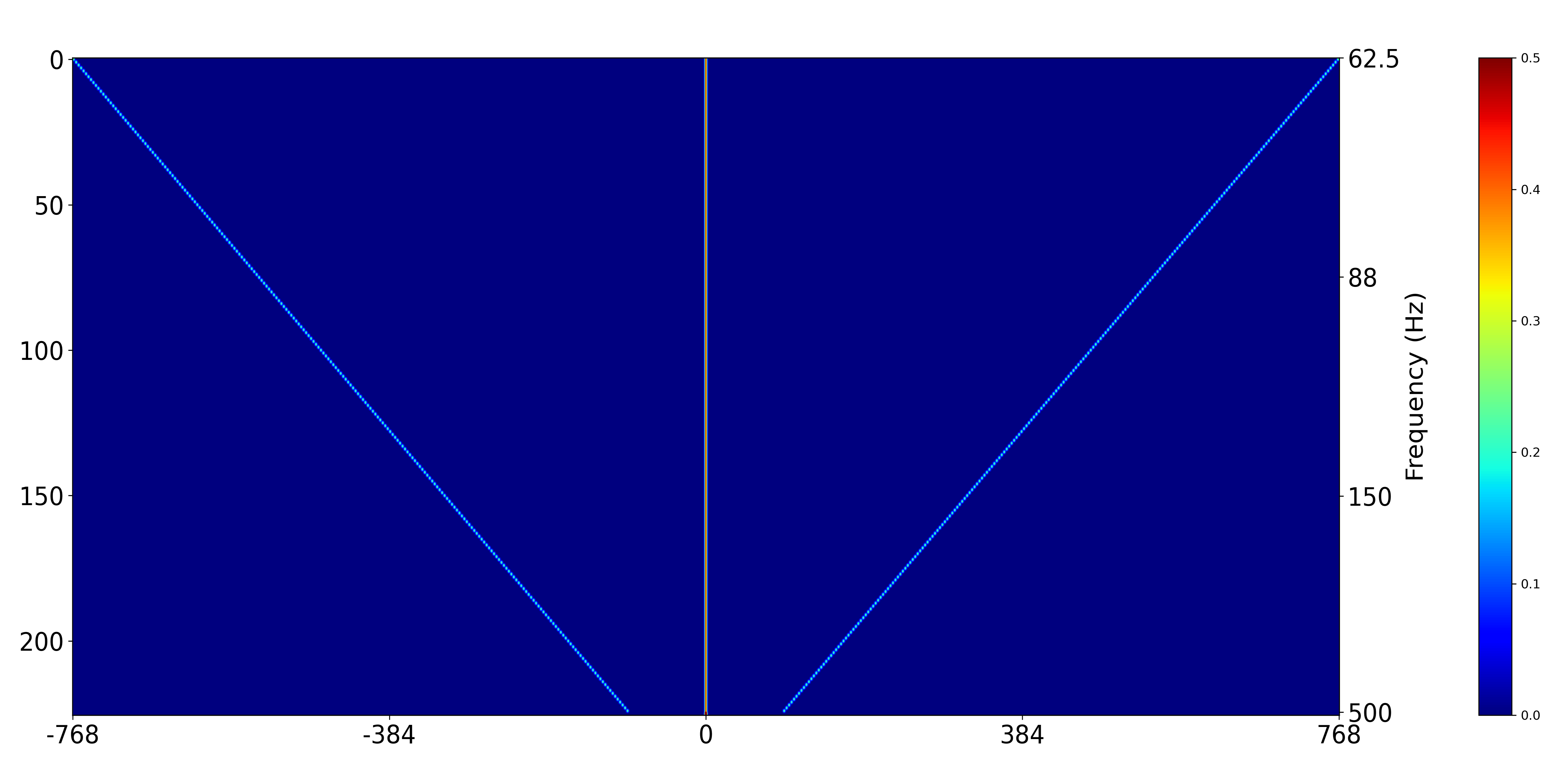}
	\caption{The visualization of the weight of the Conv2D. The vertical and horizontal axes respectively represent the channel and the kernel size.}
	\label{fig:figure2}

\end{figure}

For each candidate F0 in the target frequencies range, we design the comb filter according to Eq.~(\ref{eq2}). Note that the unvoiced frames are not processed. At the training stage, the model chooses a filter corresponding to the ideal F0 for filtering. The comb filtering algorithm is implemented by using a 2-dimensional convolutional Layer (Conv2D) to take advantage of its parallelism thus speeding up the computation. The kernel size and the stride of the Conv2D are \(\left(K,1\right)\) and \(\left(1,1\right)\), respectively. The numbers of the input and output channels are 1 and \(N+1\), respectively. As visualized in Figure~\ref{fig:figure2}, the weights can be denoted as a 4D tensor \(\mathbf{W}\in{\mathbb{R}}^{\left(N+1\right)\times 1 \times K\times1}\), where \(K=2MT_{max}+1\) depends on the longest period of the F0s, denoted as \(T_{max}=\frac{f_s}{f_{min}}\). The weight of the Conv2D is initialized using the comb filter coefficients, which can be expressed as
\begin{equation}
	\mathbf{W}\left[i,0,j,0\right] = \begin{cases}
		w_{\pm k},&{\text{if}}\ j=MT_{max} \pm kT \\ 
		{0,}&{\text{otherwise.}} 
	\end{cases}
	\label{eq5}
\end{equation}
where \(w_k\) represents the filter coefficient as in Eq.~(\ref{eq2}) and \(k\in\left[0, 1, 2, ..., M\right]\). \(T\) is the period of the \(i\)-th F0. The weight is fixed during training.

The detailed filtering processing is described below. We denote the input speech as \(x\left(t\right)\), and it is split into overlapped frames with the frame size \(N_f\) and the hop size \(N_h\), denoted as \(\mathbf{X} \in \mathbb{R}^{N_f \times N_t}\), where \(N_t\) is the number of the frames. At the same time, we split \(x(t)\) with the frame size \(N_f+2MT_{max}\) and the hop size \(N_h\) and obtain the chunked signal \(\mathbf{X}_{in} \in \mathbb{R}^{1\times\left({N_f+2MT_{max}}\right)\times N_t}\) for filtering. Note that we pad \(MT_{max}\) points on each side of the audio and \(\mathbf{X}_{in}\) have the same total frame number as \(\mathbf{X}\). The filtering process is expressed as the calculation of 2-dimensional convolution, expressed as
\begin{equation}
	\mathbf{X}_c=\text{Conv2D}\left(\mathbf{W},\mathbf{X}_{in}\right)
	\label{eq6}
\end{equation}
where \(\mathbf{X}_c\in{\mathbb{R}}^{(N+1)\times N_f \times N_t}\) is the output of the Conv2D. During training, we firstly get the one-hot tensor \(\mathbf{f}_{one-hot}\in{\mathbb{R}}^{(N+1)\times1\times N_t}\) from the ideal F0 label \(\mathbf{f}_t\). If the index of the F0 at time \(t\) is \(n\), \(\mathbf{f}_{one-hot}[n,:,t]=1\) and 0 elsewhere. The filtering result \(\mathbf{X}_{cf}\) is then calculated by the multiplication at the F0 dimension as
\begin{equation}
	\mathbf{X}_{cf}=\sum_{i}{\mathbf{X}_c[i,:,:]\mathbf{f}_{one-hot}[i,:,:]}
	\label{eq7}
\end{equation}
where \(\mathbf{X}_{cf} \in{\mathbb{R}}^{N_f \times N_t}\). In order to apply different filter strength in different bands, the noisy spectrogram \(\mathbf{Y}\) is added to the filtered spectrogram \(\mathbf{Y}_{cf}\) with the filter strength \(\mathbf{R}\):
\begin{equation}
	\mathbf{Y}_{out}= (\mathbf{R} \circ \mathbf{Y}_{cf} + (\mathbbm{1}-\mathbf{R}) \circ \mathbf{Y}) \circ \mathbf{G}
	\label{eq8}
\end{equation}
where \(\mathbf{Y}=\text{FFT}(\mathbf{X})\), \(\ \mathbf{Y}_{cf}=\text{FFT}(\mathbf{X}_{cf})\in\mathbb{C}^{{(N}_f/2+1)\times N_t}\). The interpolated gain \(\mathbf{G}\) and the filter strength \(\mathbf{R}\) both in \({\mathbb{R}}^{{(N}_f/2+1)\times N_t}\) are estimated by the DNN and \(\circ\) indicates element-wise multiplication. \(\mathbbm{1}\in{\mathbb{R}}^{{(N}_f/2+1)\times N_t}\) is a tensor of ones. To further strengthen the harmonics, we also propose a heuristic rescaling strategy to adjust the range of \(\mathbf{R}\), which is implemented by an element-wise power exponent with the coefficient \(\gamma\), expressed as
\begin{equation}
	\mathbf{Y}_{out}=( \mathbf{R}^{\gamma} \circ \mathbf{Y}_{cf} +(\mathbbm{1}-\mathbf{R}^{\gamma}) \circ \mathbf{Y})\circ \mathbf{G}
	\label{eq9}
\end{equation}
The enhanced signal in the time domain is obtained by inverse Fourier transform and overlap-add from the enhanced spectrogram \(\mathbf{Y}_{out}\).

Note that the Conv2D computes the filtering results of all the candidate F0s at the same time during training, leading to a huge amount of computational cost. However, only a little computational load is required for inference because there is only one F0 at each time step. Assuming that the period of the voiced speech at time step \(t\) is \(T\). The filtered spectrum \(\mathbf{Y}_{cf}[:,t]\) can be calculated by a frequency-domain convolution during inference, expressed as:
\begin{equation}
	\mathbf{Y}_{cf}[:,t]=\sum_{k}{w_k\left(\mathbf{X}[:,t]e^{-j\omega kT}\right)}
	\label{eq10}
\end{equation}
where \((\mathbf{X}[:,t]e^{-j\omega kT})=\text{FFT}(\mathbf{X}_{in}[0,(MT_{max}-kT):(MT_{max}+N_f-kT),t])\). Therefore, the proposed method is suitable for real-time implementation.

\begin{figure}[tbh!]
	\vspace{-0.2cm}
	\centering
	\includegraphics[width=0.5\textwidth, height=0.25\textwidth]{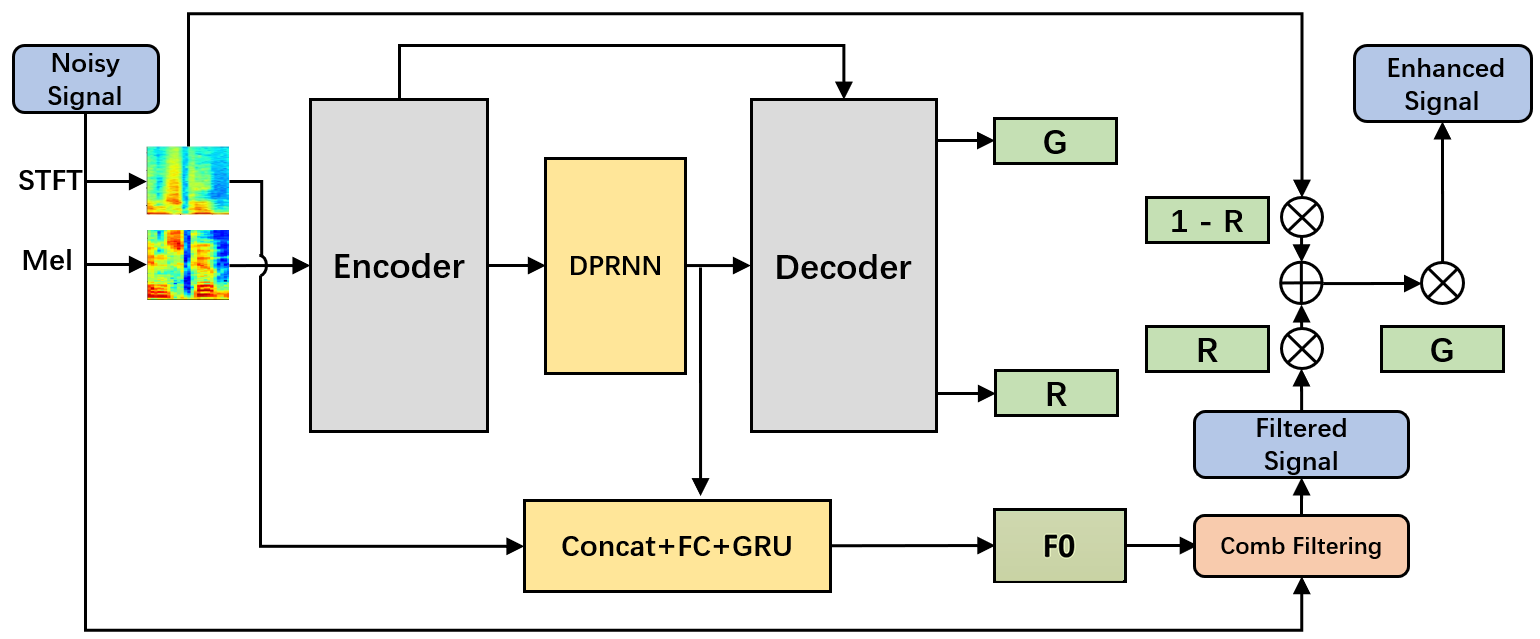}
	\caption{The structure diagram of DPCRN model.}
	\label{fig:figure3}
	\vspace{-0.3cm}
\end{figure}
\subsection{DNN model}
We use DPCRN \cite{ref14} as the core structure of the DNN model. DPCRN combines the dual-path RNN (DPRNN) \cite{ref17} and CRN \cite{ref18} in an effective way and ranked at the 3rd place in DNS-3 challenge with significantly lower computational burden than the top two models. To further reduce the computational load in its full-band implementation, we make the following modifications and the final model structure is shown in Figure~\ref{fig:figure3}.

\textbf{(a)} We use the 80-dimensional Mel-spectrogram as the input feature. The outputs are the sub-band gain (mask), sub-band filtering strength \(\mathbf{R}\) and the estimated F0. Two parallel decoders are used to get the sub-band gain and \(\mathbf{R}\) respectively, which are connected to the encoder by skip-connections.

\textbf{(b)} There are five convolutional layers in the encoder with the channels of \{32,32,32,64,64\}. The first convolutional layer has a frame look-ahead.

\textbf{(c)} The depth-separable convolution \cite{ref19} is used to reduce the computational complexity of the encoder and decoders. To alleviate the performance loss caused by the depth-separable CNNs, the gated CNN \cite{ref20} structure is added, in which an extra parallel CNN is used as a gate. With the above adjustments, the computational cost of the convolutional layers is reduced by 69\% (141 M MACs).

\textbf{(d)} The F0 estimator is trained under a multi-task learning framework. The output of the last DPRNN is compressed to 128 dimensions by a fully-connected layer. Then it is concatenated with the features of the original logarithmic spectrogram below 2 kHz. The concatenated feature is then processed by a GRU and a fully-connected layer with a sigmoid activation to obtain the final F0 estimation. 

We use the complex compressed spectrum MSE \cite{ref5} as the loss function, which is the weighted sum of the magnitude loss and the complex loss, expressed as
\vspace{-0.2cm}
\begin{equation}
	\begin{aligned}
		L_{se} = & \frac{\left(1-\lambda\right)}{2}\left(MSE_a\left(\left|S\right|^c,\left|{\hat{S}}_0\right|^c\right)+MSE_a\left(\left|S\right|^c,\left|\hat{S}\right|^c\right)\right) \\
		& + \lambda MSE\left(S^c,{\hat{S}}^c\right) 
	\end{aligned}\label{eq11}
\end{equation}
where \(S\) and \(\hat{S}\) denote the clean and the output spectrogram, \(S^c\) and \({\hat{S}}^c\) denote the complex spectrogram after magnitude compression with an exponent of \(c\) \cite{ref21}. \({\hat{S}}_0\) denotes the enhanced audio only by sub-band gains. As for the magnitude loss, the asymmetric MSE loss \cite{ref22} is used for better speech preservation, denoted as \(MSE_a(\cdot,\cdot)\). \(\lambda\) represents the weight of the magnitude loss.
The final loss function can be expressed as a weighted sum of \(L_{se}\) and \(L_{pitch}\):
\begin{equation}
	L=L_{se}+ \alpha L_{pitch} 
	\label{eq13}
\end{equation}
where \(c\), \(\lambda\) and \(\alpha\) are set to 0.3, 0.3 and 0.1 respectively.

\section{Experiments}
\subsection{Datasets and evaluation metrics}
We conduct the experiments on the DNS4 dataset \cite{ref1} and the VCTK-DEMAND \cite{ref23} dataset. For the DNS4 dataset, we select 60000 clean utterances and pre-clean them. During training, the audio is randomly clipped into 5-second segments, 20\% of which are convolved with room impulse responses (RIRs) generated by the image method to simulate the real reverberation scenarios. After adding reverberation, the audio will be mixed with noises and the signal-to-noise ratio (SNR) of the mixture is randomly sampled between -5 and 30 dB. Finally, level rescaling is applied for augmentation. As for VCTK datasets, we use the data of 84 speakers (about 27 h) for training, in which 3 hours audio is selected for validation. The DNS-4 blind test set and VCTK-DEMAND test set is used for comparison. All the audio used is sampled at \(f_s =\) 48kHz.

We use DNSMOS P.835 \cite{ref25} for evaluation, and add additional objective metrics for the VCTK test set, including SDR \cite{ref26}, PESQ \cite{ref27} and STOI \cite{ref28}.

\subsection{Parameter settings}
In our model, the frame size \(N_f\) and hop size \(N_h\) are 32 ms and 8 ms respectively. The lowest candidate F0 \(f_{min}=\) 62.5 Hz, and the longest period \(T_{max}=\)16 ms, leading to a total latency of 48 ms. The dimensions of the hidden states of DPRNN and GRU are set to 64 and 128, respectively. The comb filtering brings the computational load of \textbf{53G MACs} per sample during training and \textbf{0.53 M MACs} in inference. For the whole model, the number of parameters is \textbf{0.43 M} and the computational cost is \textbf{300M MACs}. As for rescaling strategy, \(\gamma\) is set to 0.5.

\subsection{Models for comparison}
In this paper, DPCRN denotes the model with only the decoder of sub-band gains, and DPCRN-CF is the model with the learnable comb filter. We compare the proposed models with PercepNet \cite{ref10}, DeepFilterNet \cite{ref11} and DeepFilterNet2\footnote{Hendrik Schr{\"o}ter and Alberto N. Escalante, ``Deepfilternet2: Towards real-time speech enhancement on embedded devices for full-band audio," arXiv preprint ArXiv:2205.05474, 2022.}. Taking the model structure and the latency into account, we also construct DPCRN-PN and DPCRN-DF, which are the DPCRN versions of PercepNet and DeepFilterNet. As for DPCRN-PN, the same MSE loss of the filtering strength as \cite{ref10} is utilized. In DPCRN-DF, the sub-band gains and the coefficients of DeepFiltering are adopted as the outputs from the two decoders. The order of the DeepFiltering is set to 5.

We also conducte the ablation experiments on the F0 estimator, in which pYIN algorithm is used for comparison. As for the ablation on the comb filter, we build a model with the filter weights are initialized as Figure~\ref{fig:figure2} and is learnable during training. Note that a regularization term is added to ensure the symmetry of the filter weights.
\begin{table}[!htbp]
	\setlength{\abovecaptionskip}{2pt}
	\caption{The objective results on VCTK-DEMAND test set and the computational cost of the models.}
	\setlength\tabcolsep{3pt}
	\label{tab:table1}
	\centering
	\begin{tabular}{ccccccl}
		\toprule
		Model   & SDR (dB)      & STOI      & PESQ   & MACs (M)    \\
		\midrule
		Noisy   &  8.39     &    0.921         &   1.97        & - \\
		PercepNet \cite{ref10}  & -       & -              & 2.73 & 800         \\
		DeepFilterNet \cite{ref11}     & 16.6         & 0.942      & 2.81  &   350       \\
		DeepFilterNet2\(^1\)    & -         & 0.943         & 3.08  & 360        \\
		DPCRN    & 17.9         & 0.947               & 3.03       & \textbf{299}     \\
		DPCRN-PN    & 18.3         & 0.945               & 3.06    & 300       \\
		DPCRN-PN + ideal \(\mathbf{R}\)    & \textbf{18.5}         & 0.949               & 3.10   & 300        \\
		DPCRN-DF    & 18.1         & 0.944               & 3.06         &310  \\
		DPCRN-CF    & 18.5         & \textbf{0.949}               & 3.12       & 300  \\
		DPCRN-CF + rescaling    & 18.4         & 0.948     & \textbf{3.13}        &300    \\
		\bottomrule
	\end{tabular}
\end{table}
\begin{table}[]
	\setlength{\abovecaptionskip}{2pt}
	\caption{The DNSMOS results on DNS-4 blind test set.}
	\vspace{0.1cm}
	\label{tab:table2}
	\centering
	\begin{tabular}{ccccl}
		\toprule
		Model   & SIG       & BAK      & OVL        &  \\
		\midrule
		Noisy   & 4.14      & 2.94            & 3.29          &  \\
		DeepFilterNet \cite{ref11}  & \textbf{4.14}       & 4.18              & 3.75         &  \\
		DPCRN    & 4.04         & 4.28               & 3.67          &  \\
		DPCRN-PN   & 4.05       & 4.34            & 3.73          &  \\
		DPCRN-DF & 4.04       & 4.30               & 3.70          &  \\
		DPCRN-CF & 4.10       & 4.36               & 3.77 &  \\
		DPCRN-CF+rescaling & 4.10 & \textbf{4.37} & \textbf{3.79} \\
		\bottomrule
	\end{tabular}
	\vspace{-0.5cm}
\end{table}

\subsection{Results}
The objective results on the VCTK-DEMAND test set are presented in Table~\ref{tab:table1}, where the multiply-accumulate operations per second (MACs) of the models are also presented. It can be seen that our baseline DPCRN outperform the original PercepNet and DeepFilterNet with fewer computational cost because of longer frame size and latency. Even under the same model structure and latency, DPCRN-CF exceeds DPCRN-PN and DPCRN-DF in terms of all three metrics.  DPCRN-CF also outperforms DPCRN-PN using ideal filtering strength in terms of PESQ, demonstrating that the end-to-end training paradigm gets better performance than the upper bound of the rule-based method.

In Table~\ref{tab:table2}, we compare the proposed model with others on the DNS blind test set. The learnable comb filter significantly improves the OVLMOS of the baseline DPCRN by 0.1. DPCRN-CF also exceeds all the other models, which is consistent with the results in Table~\ref{tab:table1}. Besides, compared with the original DeepFilterNet in DNS4, the proposed model shows better performance in terms of BAKMOS and OVRMOS. Furthermore, by comparing Table~\ref{tab:table1} and Table~\ref{tab:table2}, although rescaling strategy reduces objective metrics, it can further reduce the inter-harmonic noise and improve BAKMOS while maintaining SIGMOS.

\begin{figure}[tbh!]
	\centering
	\includegraphics[width=0.45\textwidth]{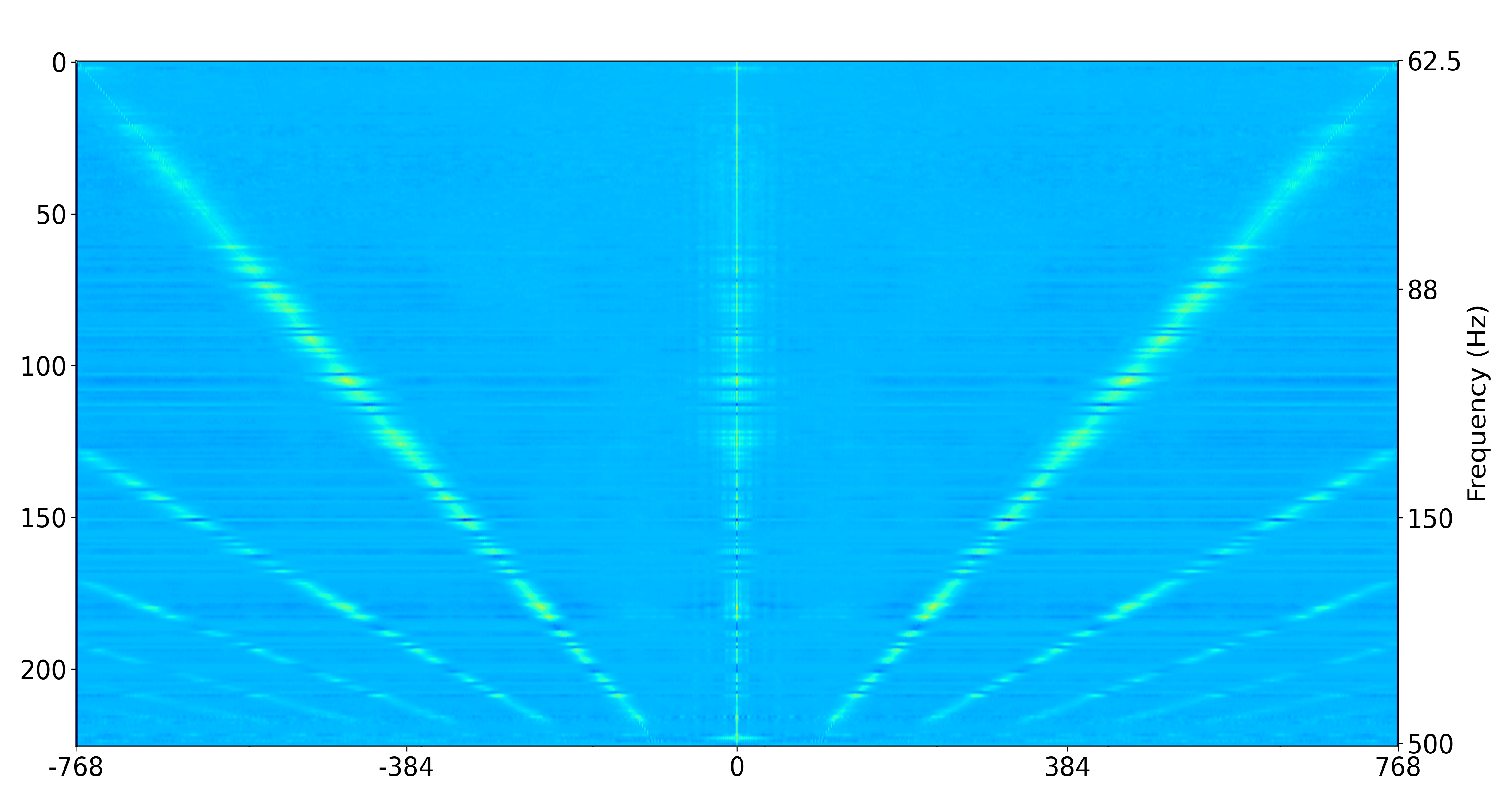}
	\caption{The visualization of the weights of the learnable comb filter.}
	\label{fig:figure4}
\end{figure}

As the results of the ablation experiments shown in Table~\ref{tab:table3}, our proposed DNN-based method achieves better performance than the rule-based YIN-based F0 estimation method. It can also be seen that the learnable filter weights bring a small improvement over the fixed weights. The visualization of the learned coefficients are shown in Figure~\ref{fig:figure4}, we can find that compared with the initial coefficients, higher order coefficients are learned, which leads to higher BAK scores. Integrating the results of the ablation study, the advantages of the end-to-end optimization paradigm are demonstrated

Although the OVLMOS is significantly improved compared to noisy samples in our experiment, the over suppression caused by the error of the F0 estimation will obviously affect the voice quality in practice. Through the bad case analysis of the enhanced audio, we find that the interference speakers are the main reason for misestimations of the F0s.
\begin{table}[tbh!]
	\setlength{\abovecaptionskip}{2pt}
	\caption{The results of the ablation experiments on the F0 estimator and learnable the comb filter.}
	\vspace{0.1cm}
	\label{tab:table3}
	\centering
	\begin{tabular}{ccccl}
		\toprule
		Model   & SIG       & BAK      & OVL        &  \\
		\midrule
		Noisy (DNS-4 blind)   & 4.14      & 2.94            & 3.29          &  \\
		\midrule
		DPCRN-CF & 4.10       & 4.36                & 3.77 &  \\
		DPCRN-CF (pYIN) & 4.09       & 4.37               & 3.75          &  \\
		DPCRN-CF (Learnable weights) & \textbf{4.12} & \textbf{4.40} & \textbf{3.80} \\
		\bottomrule
	\end{tabular}
\end{table}

\vspace{-1cm}
\section{Conclusions}

\label{sec:typestyle}
Inspired by the end-to-end training paradigm, we propose a learnable comb filter for harmonic enhancement, which is implemented by using a 2-dimensional convolutional layer. Furthermore, the end-to-end training framework is realized by combining the DNN-based fundamental frequency estimator and the comb-filtering algorithm. Experiments show that the learnable comb filter outperforms the comb filter derived from coherence function. With only 300 M MACs of computational complexity, our model achieves better results than PecepNet and DeepFilterNet in full-band SE. The personalized harmonic enhancement and more filtering forms will be explored in our future work.

\section{Acknowledgement}
This work was supported in part by the National Natural Science Foundation of China (Grants No. 12274221).

\bibliographystyle{IEEEtran}
\bibliography{mybib}

\end{document}